# Development of a Chemistry Dynamic Load Balancing Solver with Sparse Analytical Jacobian Approach for Rapid and Accurate Reactive Flow Simulations

Yinan Yang[1*], Tsukasa Hori[1], Shinya Sawada[1], Fumiteru Akamatsu[1]

*1. Department of Mechanical Engineering, Graduate School of Engineering, Osaka University, Suita, Osaka 565-0871, Japan*

# Abstract

In addressing the demands of industrial high-fidelity computation, the present study introduces a rapid and accurate customized solver developed on the OpenFOAM platform. To enhance computational efficiency, a novel integrated acceleration strategy is introduced. Initially, a sparse analytical Jacobian approach utilizing the SpeedCHEM chemistry library was implemented to increase the efficiency of the ODE solver. Subsequently, the Dynamic Load Balancing (DLB) code was employed to uniformly distribute the computational workload for chemistry among multiple processes. Further optimization was achieved through the introduction of the Open Multi-Processing (OpenMP) method to enhance parallel computing efficiency. Lastly, the Local Time Stepping (LTS) scheme was integrated to maximize the individual time step for each computational cell, resulting in a noteworthy minimum speed-up of over 31 times. The effectiveness and robustness of this customized solver were systematically validated against three distinct partially turbulent premixed flames, Sandia Flames D, E, and F. Additionally, a comparative analysis was conducted, encompassing different turbulence models, turbulent Prandtl numbers, and model constants, resulting in the recommendation of optimal numerical parameters for various conditions. The present study offers one viable solution for rapid and accurate calculations in the OpenFOAM platform, while also providing insights into the selection of turbulence models and parameters for industrial numerical simulation.

***Keywords*:** OpenFOAM; Stiff ODE solver; Load balancing; Sandia flames; Turbulence model; NO formation prediction;

# 1. Introduction

Computational fluid dynamic (CFD) analysis constitutes an indispensable and significant aspect in the development of practical ammonia combustion furnaces for pollutant reduction applications [1, 2]. The application of such a method not only reduces costs associated with the combustion furnace design process but also shortens development cycles, facilitating the rapid proliferation of clean and stable combustion technologies [3]. CFD analysis primarily focuses on two critical aspects: computational speed and accuracy, which are often interrelated. In the context of industrial application, constraints related to computational expenses and work progress usually necessitate the use of coarse meshes, along with lower-fidelity sub-models [4]. These simulations could provide valuable insights into trends during the transition of experimental conditions. Nevertheless, they fall short in providing accurate quantitative predictions of combustion characteristics and emissions, owing to the compromises made on mesh resolution. Hence, given these considerations, the development of a rapid, precise open-source CFD solver is of great significance for industrial research.

In recent years, advancements in science and technology have significantly accelerated computer processing speeds. Also, there has been a growing depth of understanding regarding ammonia, leading to the proposal of increasingly detailed reaction mechanisms for ammonia combustion [5, 6]. This development holds promising prospects for the realization of high-fidelity reaction flow simulations. However, it is worth noting that performing computations with such detailed reaction mechanisms can still incur substantial expenses. Because in finite-rate chemistry calculations, for each computational cell, it is necessary to solve differential equations for the evolution of individual species in addition to solving the Navier-Stokes equations for momentum and energy. This computational load escalates with the increasing number of reactive species, resulting in higher computational costs and significant storage requirements [7]. Therefore, even within the realm of industrial applications employing Reynolds-averaged Navier–Stokes (RANS) simulations, the incorporation of detailed reaction mechanisms remains prohibitively costly.

Generally speaking, chemistry evaluation comprises the most computationally demanding part of the simulations. In finite-rate chemistry simulations that employ detailed reaction mechanisms, the substantial computational expense primarily stems from three key factors, they are, the size of the reaction mechanism, the load imbalance issue in multi-processor applications, and the grid dimensions. First of all, as mentioned previously, the use of detailed reaction mechanisms increases the number of advection equations, diffusion coefficient calculations, and therefore the stiffness of chemical-reaction Ordinary Differential Equations (ODEs). The computational cost of solving the associated stiff ODEs usually scales quadratically with the number of involved species [8]. This complexity poses significant challenges to the widespread application of detailed reaction mechanisms. The secondary issue lies in the imbalanced computational load distribution in parallel calculations. In most CFD software, parallelization is achieved through geometric domain decomposition. However, during combustion reactions, values of the thermochemical state vector among various meshes change over time, leading to explicit load imbalances among different sub-domains [9].

Lastly, the resolution of the computational grid also affects computational speed. In computational modeling, each grid is modeled as an independent homogeneous gas-phase reactor, meaning that the number of grid cells is directly proportional to the computational workload. Although most CFD software packages implement parallel computing strategy to partition the computational domain into multiple sub-domains, excessive domain partitioning can lead to communication bottlenecks among processors, ultimately diminishing the overall efficiency of the solver [10]. Moreover, in industrial combustion furnace simulations, grid refinement is necessary for some specific locations, for example, the inlet nozzle and main reaction area, to accurately capture the flow characteristics and species distribution. However, since global time steps are determined by the smallest cell size, constrained by these small cells, the global time step for reactions may become exceedingly low. This significantly increases reaction convergence time, thereby reducing computational efficiency.

Over the past three decades, researchers have proposed various computational acceleration strategies to address the aforementioned issues and have made substantial improvements. The majority of current studies focus on improving specific aspects such as reaction mechanisms [11, 12], ODE solver optimization [13–15], and addressing load imbalances in parallel computing [4, 16]. In practical applications, it is often necessary to combine these acceleration strategies to achieve a better increase in computational speed. However, the integration of multiple acceleration strategies into a customized solver may not always yield maximal speed enhancements and can sometimes have counterproductive effects, potentially impeding computational efficiency. The impact of different integrated strategies on the enhancement of computational accuracy, speed, and stability remains uncertain. Therefore, it is of great significance to explore a robust and efficient integrated acceleration strategy. The present study aims to consolidate multiple efficient acceleration strategies within the framework of the OpenFOAM platform, ultimately achieving both high-speed and stable computations. First, to enhance the performance of the ODE solver, a sparse analytical Jacobian approach utilizing the SpeedCHEM chemistry library was implemented. This significantly reduces the CPU time required for time integration of species evolution due to chemical reactions [13]; Second, the dynamic load balancing (DLB) code was employed to evenly redistribute the computational load for chemistry among multiple processes, thereby mitigating load imbalance issues in multi-processor applications [16]. Third, the Open Multi-Processing (OpenMP) method was introduced to enhance parallel computing efficiency by utilizing multi-threading, thereby mitigating the communication bottlenecks among processors [10]. Fourth, a local time-stepping (LTS) scheme, allowing for individual time steps for each cell based on the local Courant-Friedrichs-Lewy (CFL) number was adopted, thus further enhancing computational efficiency.

Following the implementation of the integrated speed-up strategy, a crucial aspect involves validating the predictive accuracy of the customized solvers through benchmark experiments. Within the context of industrial ammonia combustion, the present study aims to identify a benchmark experiment that is not only characterized by accuracy and reliability, but also encompasses extensive measurements of critical parameters such as the temperature field, velocity field, and combustion reactants, with a specific focus on NO distribution. For this purpose, the Sandia flames D, E, and F (hereafter abbreviated as flames D-F) were selected as the validation [17]. This type of turbulent partially premixed methane/air flame has been widely investigated and is seen as essential for various studies [18, 19].

To enhance the predictive accuracy of our customized solver, a series of optimizations have also been applied to the reactingFoam solver within the OpenFOAM platform. These optimizations involve modifications to the governing equations, turbulence models, combustion models, and radiation models, as elaborated upon in Section 2. Furthermore, it is worth noting that, to the best of our knowledge, certain fundamental questions on the selection of model parameters have not been well addressed. For example, selecting an appropriate turbulence model is a crucial problem for RANS modeling. However, currently, there is no consensus on which turbulence models provide the most accurate predictions of temperature and velocity fields, as well as species distribution. Furthermore, it is unclear which turbulence model is universally applicable across various experimental conditions and how to adjust the turbulence Prandtl number when the jet inlet velocity is changeable. Another key focus of industrial application research is to minimize the generation of harmful species. Nonetheless, the accuracy of predicting NO emissions with different turbulence models, particularly for flames E and F, which have yet to be comprehensively studied, remains uncertain. These questions form the basis that motivates the validation part.

In summary, the paper is structured as follows. First, Section 2 elaborates on the implementation specifics of the customized solver and evaluates the integrated acceleration strategy. Section 3 proceeds with an evaluation of the predictive accuracy of the customized solver, utilizing the flames D-F as the benchmark experiment. Furthermore, it provides suggestions for the selection of various turbulence models, turbulent Prandtl numbers, and model constants based on the customized solver. The overall conclusions are summarized in Section 4.

## 2. Computational modeling

### 2.1 Governing equations

In this study, the Sandia flame experiment was reproduced by employing RANS equations. Favre averages were utilized to account for variable-density effects. The transport equations of mass, momentum, chemical species, and enthalpy are respectively described by Eqs. (1) - (4)

$$\frac{\partial(\bar{\rho})}{\partial t} + \nabla \cdot (\bar{\rho}\tilde{u}) = 0 \tag{1}$$

$$\frac{\partial(\bar{\rho}\tilde{u})}{\partial t} + \nabla \cdot (\bar{\rho}\tilde{u}\tilde{u}) = -\nabla \bar{p} + \nabla \cdot (\bar{\tau}) + \bar{\rho}g \tag{2}$$

$$\frac{\partial(\bar{\rho}\tilde{Y}_s)}{\partial t} + \nabla \cdot (\bar{\rho}\tilde{u}\tilde{Y}_s) = \nabla \cdot (\bar{\rho}D_{eff}\nabla\tilde{Y}_s) + \bar{R}_s \tag{3}$$

$$\frac{\partial(\bar{\rho}\tilde{h})}{\partial t} + \nabla \cdot (\bar{\rho}\tilde{u}\tilde{h}) + \frac{\partial(\bar{\rho}\tilde{E})}{\partial t} + \nabla \cdot (\bar{\rho}\tilde{u}\tilde{E}) = \frac{\partial \bar{p}}{\partial t} + \nabla \cdot (\alpha_{eff}\nabla\tilde{h}) + \bar{Q}_c + \bar{Q}_r \tag{4}$$

where the overbar and tilde represent the Reynolds and Favre averages, respectively; $\rho$ is the density and $u$ represent the vector unit of fluid velocity. The variables $p$ and $g$ denote the pressure in every point of the fluid and gravity acceleration; $h$ is the specific sensible enthalpy and $E$ denotes the total energy; $D_{eff}$ and $\alpha_{eff}$ represent the effective diffusion coefficient and the effective thermal diffusion coefficient, respectively. Moreover, Eq. (2) calculates $\tau$, the stress tensor, whose variable presents different expressions in various turbulence models [20]. The chemical species equation involves $Y_s$ and $R_s$, referring to the mass fraction and reaction rate of species *s*, respectively. The Eddy Dissipation Concept (EDC) is applied as closure for the source term. Additionally, the Partially Stirred Reactor (PaSR) model is also used for prediction and comparison, as discussed in Section 3.4. Heat source terms $Q_c$ and $Q_r$ are due to combustion and thermal radiation, respectively.

**2.2 Turbulence model**

From the momentum equations, certain source terms cannot be solved directly, thereby necessitating the use of a turbulence model to depict the flow field with a reasonable computational cost. In the present study, the selected turbulence model includes the Standard *k-ε* model (KE model), the Re-Normalization Group *k-ε* model (RngKE model), and one variant of the Reynolds stress model (LRR model).

2.2.1 Standard *k-ε* and Re-Normalization Group *k–ε* model

The KE model serves as a reference point for comparing prediction results. This simple semi-empirical model is based on the eddy viscosity concept. Numerically, the KE model involves solving two partial equations for turbulent kinetic energy *k* and turbulence eddy dissipation *ε*, as described in Eqs. (5) and (6), respectively.

$$\frac{\partial(\bar{\rho}\tilde{k})}{\partial t} + \nabla \cdot (\bar{\rho}\tilde{u}\tilde{k}) = \nabla((\mu + \frac{\mu_{eff}}{\sigma_k})\nabla\tilde{k}) + P_k - \bar{\rho}\tilde{\varepsilon} + P_{kb} - Y_M \quad (5)$$

$$\frac{\partial(\bar{\rho}\tilde{\varepsilon})}{\partial t} + \nabla \cdot (\bar{\rho}\tilde{u}\tilde{\varepsilon}) = \nabla((\mu + \frac{\mu_{eff}}{\sigma_\varepsilon})\nabla\tilde{\varepsilon}) + \frac{\tilde{\varepsilon}}{\tilde{k}}(C_{\varepsilon 1}P_k - C_{\varepsilon 2}\bar{\rho}\tilde{\varepsilon} + C_{\varepsilon 1}P_{\varepsilon b}) \quad (6)$$

where $P_k$ is the turbulence kinetic energy production due to viscous forces; $P_{kb}$ and $P_{\varepsilon b}$ represent the influence of the buoyancy forces; $Y_M$ denotes the contribution of the fluctuating dilatation in compressible turbulence to the overall dissipation rate; $\mu_{eff}$ is the effective dynamic eddy viscosity linked to the turbulence kinetic energy and turbulence eddy dissipation, is calculated as follows:

$$\mu_{eff} = \bar{\rho} C_\mu \frac{\tilde{k}^2}{\tilde{\varepsilon}} \qquad (7)$$

The KE model includes five adjustable model constants with recommended values given by Ref. [21], namely, $C_{\varepsilon1}$ = 1.44, $C_{\varepsilon2}$ = 1.92, $C_\mu$ = 0.09, $\sigma_k$ = 1.0, and $\sigma_\varepsilon$ = 1.3.

The RngKE model serves as an advanced version of the KE model, characterized by an additional term in its $\varepsilon$ equation that improves its accuracy for rapidly strained flows [22]. Although the RngKE model is generally regarded as more reliable and precise than the KE model, it has not yet gained widespread adoption in practical applications. The equations of the RngKE model are expressed as follows:

$$\frac{\partial(\bar{\rho}\tilde{k})}{\partial t} + \nabla \cdot (\bar{\rho}\tilde{u}\tilde{k}) = \nabla((\mu + \frac{\mu_{eff}}{\sigma_k})\nabla\tilde{k}) + P_k - \bar{\rho}\tilde{\varepsilon} + P_{kb} - Y_M \qquad (8)$$

$$\frac{\partial(\bar{\rho}\tilde{\varepsilon})}{\partial t} + \nabla \cdot (\bar{\rho}\tilde{u}\tilde{\varepsilon})$$
$$= \nabla((\mu + \frac{\mu_{eff}}{\sigma_\varepsilon})\nabla\tilde{\varepsilon}) + \frac{\tilde{\varepsilon}}{\tilde{k}}(C_{\varepsilon1}P_k - C_{\varepsilon2}\bar{\rho}\tilde{\varepsilon} + C_{\varepsilon1}P_{\varepsilon b}) - R_\varepsilon \qquad (9)$$

where the main difference between the RngKE model and KE model lies in the additional term $R_\varepsilon$ in the $\varepsilon$ equation and is given by Eq. (10)

$$R_\varepsilon = \frac{C_\mu \bar{\rho} \eta^3 (1 - \eta/\eta_0)}{1 + \beta\eta^3} \frac{\varepsilon^2}{k} \qquad (10)$$

where $\eta \equiv S_k/\varepsilon$, $\eta_0$ = 4.38, $\beta$ = 0.012.

The model constants for the RngKE model used by default and are given as $C_{\varepsilon1}$ = 1.42, $C_{\varepsilon2}$ = 1.68, $C_\mu$ = 0.0845.

### 2.2.2 Reynolds stress equation model

The Reynolds Stress model (RSM) is a classic turbulence model consisting of three variants. In this study, the LRR-IP model (hereafter abbreviated as LRR model), developed by Launder et al. [23], was applied. Unlike the KE model, the LRR model abandons the isotropic eddy-viscosity hypothesis and solves transport equations for the Reynolds stresses, along with an equation for the dissipation rate, to close the Reynolds-averaged Navier-Stokes equations. Therefore, in 2-D flows, five additional transport equations are required compared to seven in 3-D flows. The equations of the LRR model are expressed as follows:

$$\frac{\partial(\overline{\rho u_i u_j})}{\partial t} + \frac{\partial(\bar{\rho}\tilde{u}_k \overline{u_i u_j})}{\partial x_k} = P_{ij} - \frac{2}{3}\delta_{ij}\bar{\rho}\tilde{\varepsilon} + \phi_{ij} + P_{ij,b} + \frac{\partial}{\partial x_k}\left((\mu + \frac{2}{3}C_s\bar{\rho}\frac{\tilde{k}^2}{\tilde{\varepsilon}})\frac{\partial \overline{u_i u_j}}{\partial x_k}\right) \quad (11)$$

$$\frac{\partial(\bar{\rho}\tilde{\varepsilon})}{\partial t} + \frac{\partial(\bar{\rho}\tilde{u}_k \tilde{\varepsilon})}{\partial x_k} = \frac{\partial}{\partial x_k}\left((\mu + \frac{\mu_{eff}}{\sigma_{\varepsilon Rs}})\frac{\partial \tilde{\varepsilon}}{\partial x_k}\right) + \frac{\tilde{\varepsilon}}{\tilde{k}}(C_{\varepsilon 1}P_k - C_{\varepsilon 2}\bar{\rho}\tilde{\varepsilon} + C_{\varepsilon 1}P_{\varepsilon b}) \quad (12)$$

where $\overline{u_i u_j}$ are the Reynolds stresses; $P_{ij,b}$ and $P_{\varepsilon b}$ denote the production due to the buoyancy; $P_{ij}$ is the exact production term. $\Phi_{ij}$ is the pressure-strain correlation influencing the Reynolds stresses redistribution. Detailed expressions of mentioned turbulence models can be found in Ref. [23].

The default values of the model constants $C_{\varepsilon 1}$ and $C_{\varepsilon 2}$ are 1.44 and 1.92, respectively. Notably, the value of $C_{\varepsilon 1}$ is reported in the literature within the range of 1.44 – 1.5. In the present study, $C_{\varepsilon 1}$ is adjusted to 1.48 and its influence is discussed, with an ultimate aim of identifying the optimal value for accurate prediction.

**2.3 Radiation model**

Radiative heat transfer is a crucial factor in turbulent combustion systems, serving as the primary energy transfer mechanism in high-temperature devices. Recent research by Tessé et al. [24] has demonstrated that turbulence-radiation interactions can significantly increase the radiative heat loss, accounting for approximately 30% of the total chemical heat release. Radiative heat transfer directly affects the temperature field, consequently affecting chemical kinetics, most notably the formation of NO, which exhibits high sensitivity to temperature variations. Therefore, implementing an accurate radiation model is essential, especially in the context of industrial applications.

In the current investigation, we utilized a combined radiation model comprising the Discrete Ordinates Method (DOM) and the Weighted Sum of Gray Gases (WSGG) absorption emission model. For gaseous mixtures, scattering is typically negligible, hence the Radiative Transfer Equation (RTE) adopted in this study only accounts for the radiation attenuation and augmentation due to absorption and emission, respectively. The simplified RTE equation is formulated in Eq. (13)

$$\frac{dI_\eta}{dS} = -\kappa_\eta I_\eta + \kappa_\eta I_{b\eta} \tag{13}$$

where, $I_\eta$ is the spectral radiation intensity at wavenumber $\eta$ and along the path length $s$; $I_{b\eta}$ is the Planck blackbody spectral radiation intensity, and $\kappa_\eta$ is the spectral absorption coefficient of the medium.

The WSGG model is a global model in which the absorption spectrum of a given species or mixture of species is represented by a set of $j$ gray gases and transparent windows [25]. Each gray gas $j$ has a unique absorption coefficient $\kappa_j$ and is assumed to occupy a fixed, yet mostly non-contiguous, portion of the spectrum. Thus, the RTE for the gas $j$ takes the form

$$\frac{dI_j}{dS} = -\kappa_j I_j + \kappa_j \alpha_j I_b \tag{14}$$

where $I_j$ represents the local intensity associated with gray gas $j$, while $I_b$ represents the total blackbody radiation intensity related to the local temperature; $\alpha_j$ is an emission weighted factors representing the fraction of blackbody energy that lies within the portion of the radiation spectrum occupied by gas $j$ [25].

Once Eq. (14) has been solved for all gray gases $j$, the local total intensity $I$ can be determined by summing up the contribution of each gray gas partial intensity, including the partial intensity related to the transparent windows. Specifically, $I = \sum_{j=0}^{j} I_j$. For a more detailed understanding of the DOM coupled with the WSGG method, readers are recommended to refer to the work by Modest [26].

**2.4 Computational domain details**

As mentioned in Section 1, Sandia flames D, E, and F were selected in the present study to validate the accuracy of our customized solver. Sandia flame is a partially turbulent premixed flame where the fuel jet is surrounded by a high-temperature and diluted co-flow (pilot zone). The velocity field and scalar data were measured by the University of Darmstadt and the Sandia National Laboratory, respectively [17]. The calculations obtained will be compared to the reference data mentioned above.

Figure 1 depicts the schematic of the Sandia flame burner. The burners are identical in construction for flames D-F, except for the differences in the jet and pilot inlet velocity. Specifically, the burner is composed of a jet nozzle with a diameter of $D$ = 7.2 mm, surrounded by a wide pilot nozzle with inner and outer diameters of 7.7 mm and 18.2 mm, respectively. In the present study, pre-inlet nozzles for both the jet and pilot were taken into consideration in the computational domain to obtain a fully developed turbulent velocity profile at the burner inlet where its length extends up to approximately 15 $D$. According to the recommended model dimension from Ref. [19], the axial and radial dimensions of the computational domain after the inlet was set to 76.5 $D$ and 20.83 $D$, respectively. Calculations were conducted in a 2-D domain as a sector of 5°. The lower dimensionality simulations were performed with a mesh having a thickness of one cell.

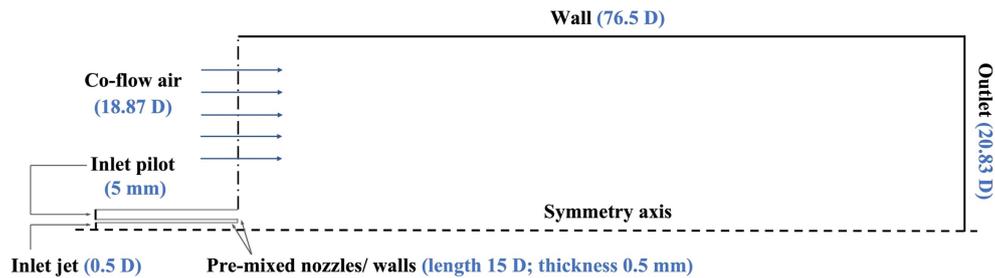

**Figure 1 Schematic of Sandia Flames D, E, and F ($D$ = 7.2 mm)**

**2.5 Reaction mechanism and integrated acceleration strategy**

To facilitate the future studies pertaining to methane-ammonia combustion simulations, we conducted Sandia flame calculations utilizing the Okafor detailed reaction mechanism. This particular mechanism was developed by integrating both the GRI-Mech 3.0 [27] and the ammonia-methane-related mechanism proposed by Tian et al. [5] and has been rigorously evaluated against experimental data [6]. Those turbulent combustion models incorporating the Okafor detailed reaction mechanism are applicable for methane-ammonia-air combustion. The detailed mechanism comprises 59 species and 356 elementary reactions.

As outlined in the introduction part, when dealing with finite-rate chemistry problems, the computational challenges predominantly stem from three factors: the size of the reaction mechanism, the load imbalance issue in multi-processor applications, and the grid dimensions. The new developments within the present study aim to integrate several efficient acceleration strategies to address these challenges. First, we introduced a chemical kinetics library called SpeedCHEM, which interfaces with the ODE system solver. This algorithm, initially postulated by Perini et al. [13], is written in the modern Fortran language. It attains high computational efficiency by evaluating functions pertinent to the chemistry ODE solver. This is achieved through the employment of optimal-degree interpolation of costly thermodynamic functions, internal sparse algebra management of mechanism-related quantities, and sparse analytical formulation of the Jacobian matrix. Without compromising computational precision, this methodology could achieve a substantial speed augmentation, notably pronounced for small to medium-sized mechanisms ($50 \leq n_s \leq 500$). Second, we employed the dynamic load balancing (DLB) code to address the load imbalance issues in multi-processor applications. The DLB framework has been extensively discussed in previous literature, manifesting numerous diverse variants [4, 16]. In the current investigation, our reference point is the DLBFoam solver recently developed by Tekgül et al. [16], implemented on the OpenFOAM platform. This algorithm harnesses MPI (Message Passing Interface) routines to redistribute the chemistry computational load evenly between processes during the simulation. Additionally, DLBFoam introduces a zonal reference cell mapping approach, which contributes to a further reduction in computational costs by mapping the chemistry solution from a reference cell rather than explicitly solving it for regions with low reactivity [9]. Third, we introduced the OpenMP technique as a means to augment the efficiency of parallel execution through multi-threading [10]. The application of the OpenMP method carries the potential to significantly diminish the need for excessive domain partitioning in high-performance computing scenarios, thereby mitigating the likelihood of encountering communication bottlenecks among processors. Fourth, we integrated a local time-stepping (LTS) scheme into our computational framework. This approach maximizes the individual time step for each computational cell based on the local CFL number, resulting in a substantial reduction in the overall computational runtime. Figure 2 illustrates the integrated acceleration strategies applied by our customized solver.

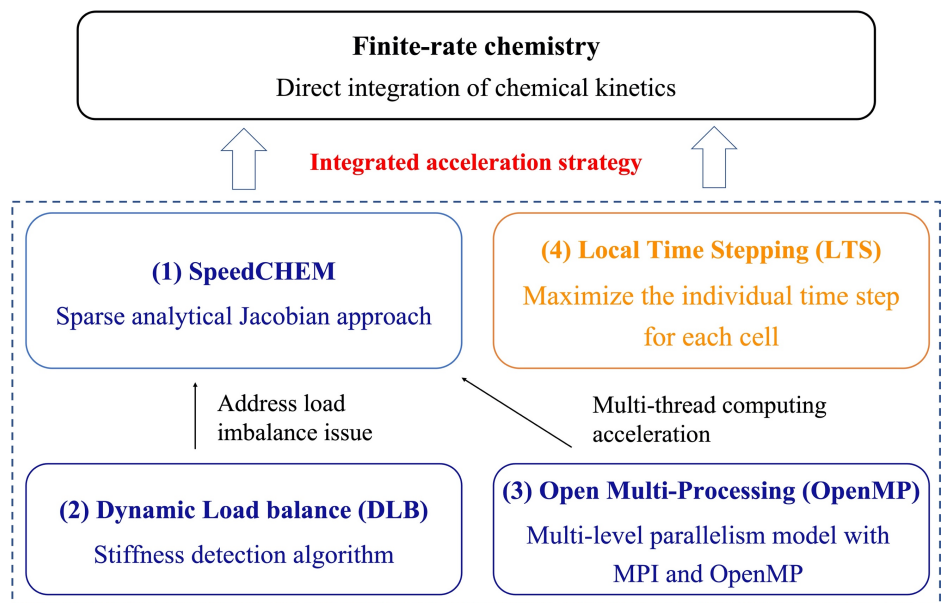

Figure 2 Schematic diagram of the integrated acceleration strategy

To provide a clear demonstration of the effectiveness of our acceleration strategies, Fig. 3 illustrates the computational speed-up achieved when calculating flame D with our customized solver. The calculation domain is decomposed into 16 processors, and all speed-up tests are conducted for 200 constant time steps of $10^{-6}$ s after ignition. Examination of the figure reveals that applying the standard ODE solver consumes considerable time, approximately 1500 s. The implementation of the DLB code yields a modest speed-up factor of around 1.4, a progression not remarkably significant when compared to prior research efforts [16]. This might be attributed to the relatively lower count meshes and the limited partitioning of computational domains. Consequently, a more substantial speed-up can be anticipated when applying DLB code in 3-D calculations [9]. Application of the SpeedCHEM + DLB approach leads to a notable speed-up by a factor of 3.34, affirming the robustness of such a chemistry solver. Finally, the integration of the OpenMP method yields substantial acceleration, with our computational strategy achieving a speed increase of nearly 31-fold compared to the standard conditions. Applying the OpenMP method enables the allocation of more threads for calculations within individual computational subdomains, which efficiently reduces communication bottlenecks in parallel calculation [10]. It is worth noting that because our customized solver employs the LTS method, that is to say, all four sets of case comparisons are based on such a method. Therefore, the integrated acceleration strategy can anticipate a higher level of speed improvement compared to the original reactingFoam solvers. Typically, achieving a converged state requires approximately a computation time of approximately 8 h, and utilizing 8 nodes in the Oakbridge-CX Supercomputer at the University of Tokyo.

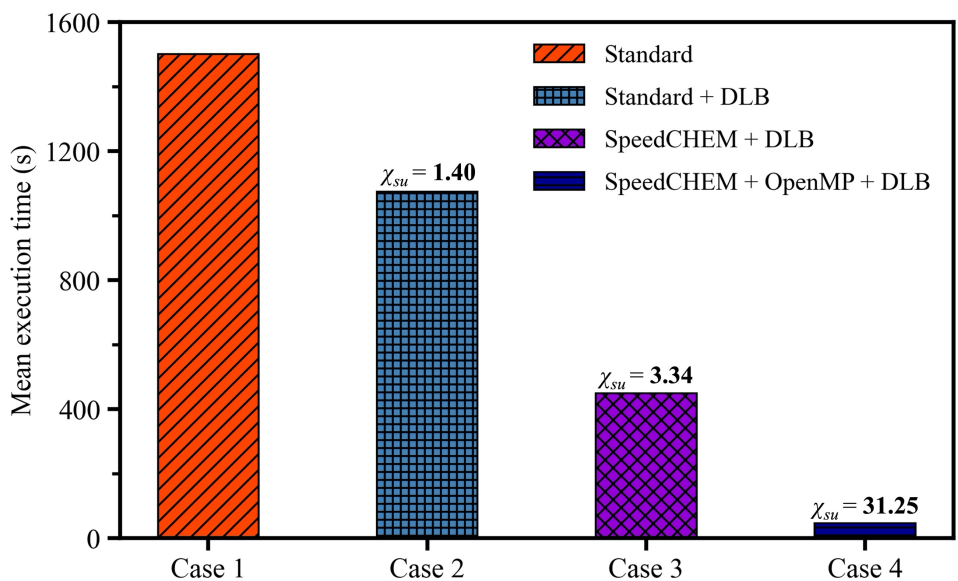

**Figure 3 Mean execution times over 200 iterations for the flame D simulation under 16 computational sub-domains ($\chi_{su}$ is the speed-up factor)**

**2.6 Numerical settings**

The detailed numerical simulation setup employing the OpenFOAM source code can be found in Table 1. A grid independence study was performed in advance with the refinement ratio between the meshes equal to 1.5. The present study calibrated the turbulent Prandtl number for different flame types (D-F) and turbulence models. Notably, the optimal range for the turbulent Prandtl number in non-isothermal circular jets has been reported to be within the interval of 0.7 to 1 in previous literature [28]. For flame D, the recommended value of 0.82, as given by Chua and Antonia [29], was used in the calculation of KE and RngKE models.

Table 1 Numerical settings for the simulation

| | Sandia Flame D | Sandia Flame E | Sandia Flame F |
|---|---|---|---|
| **Turbulence model** | Standard k-epsilon (KE) | | |
| | Re-Normalization Group k-epsilon (RngKE) | | |
| | Launder-Reece-Rodi (LRR) | | |
| **Combustion model** | Eddy Dissipation Concept (EDC) | | |
| **Radiation model** | Discrete Ordinates Method (DOM) and the Weighted Sum of Gray Gases (WSGG) absorption emission model | | |
| **ODE solver optimization** | SpeedCHEM chemistry solver coupled with Dynamic load balancing (DLB) code | | |
| **Parallel computing optimization** | Open Multi-Processing (OpenMP) method | | |
| **Computational time optimization** | Local time-stepping (LTS) scheme | | |
| **Reaction mechanism** | Okafor detailed reaction mechanism | | |
| **Mesh number** | ~ 29000 (2-D calculation) | | |
| **Discretization schemes** | 2nd order | | |
| **Turbulent Schmidt number** | 0.7 | 0.7 | 0.7 |
| **Turbulent Prandtl number** | 0.72 (LRR) | 0.82 (LRR) | 1 |
| | 0.82 (KE, RngKE) | 0.92 (KE, RngKE) | |
| **Reynolds number** | 22400 | 33400 | 44800 |
| **Velocity in the jet zone** | 49.6 m/s | 74.4 m/s | 99.2 m/s |

However, due to the significant increase in inlet jet velocity of flames E and F compared to flame D, which enhances the momentum eddy diffusivity, it becomes necessary to increase the turbulent Prandtl number during the transition from flame D to F to control the relationship between momentum and heat diffusion eddy diffusivity. Additionally, it was found that decreasing the turbulence Prandtl number by 0.1 for flames D and E, as compared to the values employed in the other two turbulence models, can yield more accurate predictions when adopting the LRR turbulence model. Detailed values concerning the utilization of turbulent Prandtl numbers in different types of flames and turbulence models are presented in Table 1. The influence of turbulent Prandtl numbers on the predictions will be discussed in Section 3.4, with the suffix (_R), which is used to indicate the optimized turbulent Prandtl number calculation cases.

## 3. Results and discussion

### 3.1 Inlet boundary conditions

When evaluating the accuracy of a customized solver, it is common practice to compare the predicted values with the experimental values. Generally, the employed combustion models, turbulence models, or reaction mechanisms are held responsible for any discrepancies between the computed and experimental values. However, Lewandowski et al. [30] noted that other factors also contribute to these discrepancies. For instance, the accuracy of the predicted results is also influenced by the inlet boundary conditions. Therefore, it is crucial to carefully consider all the relevant factors when interpreting the results of a numerical simulation.

In the present study, to ensure a fully developed turbulent velocity profile at the burner inlet, pre-inlet nozzles with an approximate length of 15 $D$ for the jet and pilot were incorporated in the computational domain. However, the presence of pre-inlet walls significantly affects turbulent flows, with regions affected by viscosity experiencing alterations in the flow. When translated to CFD calculations, differences in the inlet wall function type can influence the velocity field predictions, resulting in deviations from the actual results. Consequently, a precise near-wall region representation is critical for accurately predicting wall-bounded turbulent flows. To assist the researchers in validating their prediction results, the Sandia National Laboratories' official documentation provided correction velocity profiles at the burner inlet of flame D.

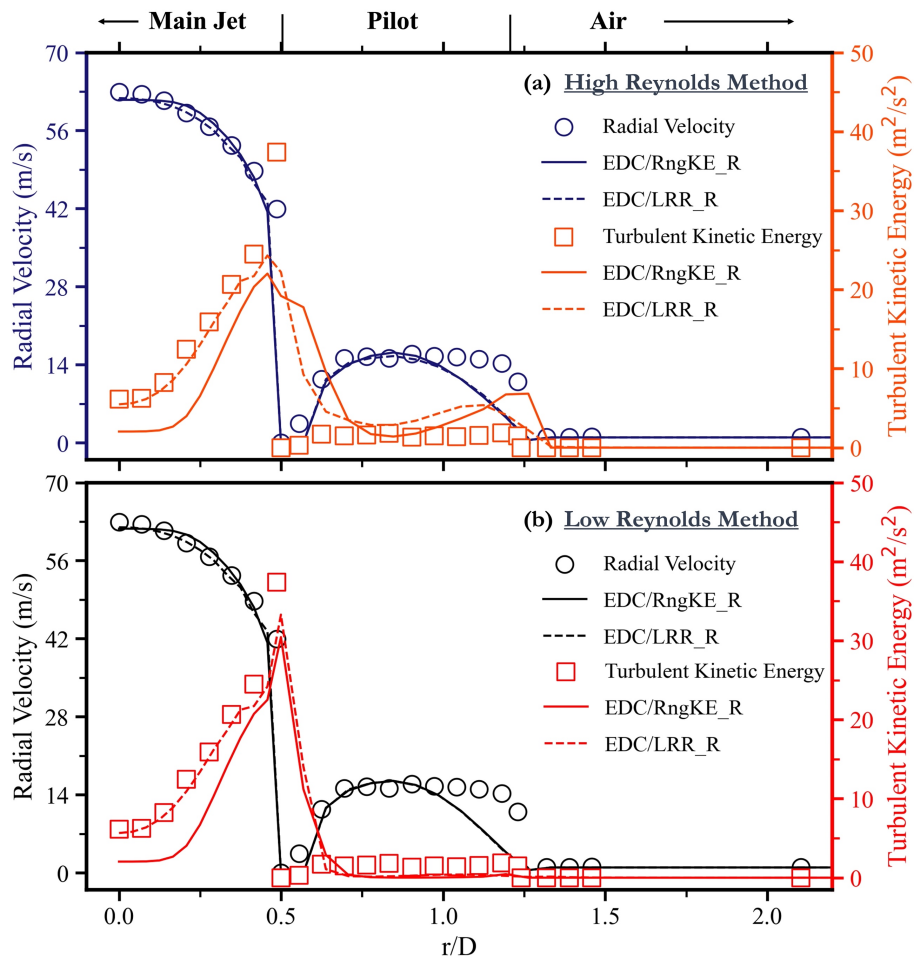

**Figure 4 Comparison of experimental and predicted velocity profiles using (a) high and (b) low Reynolds methods for flame D at the inlet boundary ($X/D = 0$)**

The examination of the $y+$ values in flame D reveals that the flow near the pre-inlet wall is not fully turbulent ($y^+ > 30$). Consequently, based on the OpenFOAM platform, the Low Reynolds (LowRe) method, which can handle a flow that is not fully turbulent, yields better predictions compared to the high Reynolds number wall functions. Figure 4 provides a comparison of experimental and predicted results utilizing these two different types of wall functions. Notably, the implementation of the LowRe method led to significant improvements in the velocity fields, particularly the turbulent kinetic energy at the burner inlet. Although some discrepancies with experimental results still exist, utilizing the LowRe method in the wall function enhanced the predictive capability for the flame D.

**3.2 Turbulence model comparison**

3.2.1 Central axis prediction

Figures 5-7 present the centerline profiles of temperature, mean mixture value, velocity, turbulent kinetic energy, major species ($CO_2$, $O_2$, $CH_4$, $N_2$), and minor species (CO, NO) for flames D-F, respectively. Notably, the KE model was not used to calculate flame F through our customized solver due to its quenched state during the calculation. In terms of temperature field prediction, all the three turbulence models exhibit a minor overestimation of the peak value along the central axis. The peak temperature differences between the RngKE and KE models were 66 K and 68 K for flame D, and 104 K and 115 K for flame E. In contrast, the peak temperature difference for flames D and E employing the LRR model is closer to the experimental value at roughly 22 K and 78 K, respectively. Concerning the flame structure, the RngKE model closely aligns

with experimental values in the flame developing and combustion region ($X/D < 45$), whereas the LRR and KE models overestimate the temperature. Nonetheless, as the flame develops, an underestimation of temperature near the flame front is observed when employing the RngKE and KE models. Collectively, good agreements with the experimental values can be observed based on the RngKE turbulence model. Reasonable temperature field predictions lead to good predictions of the species distribution. As shown in Fig. 6, the RngKE model offers higher accuracy in predicting major species than the other two models.

Regarding the velocity field predictions for flames D and E, the RngKE model exhibits significant discrepancies in the turbulent kinetic energy values despite capturing some experimental trends. Conversely, the LRR model provides superior predictions for turbulent flow owing to its ability to solve transport equations for each component of the Reynolds stress tensor, and accounts for the history and anisotropy of turbulence [20]. Consequently, the predictions generated by the LRR model are more closely aligned with the actual experimental values.

Nonetheless, the prediction accuracy for flame F diminishes when employing the RngKE and LRR turbulence models, particularly in the velocity field. Given that flame F has significant local extinction, the severity of local extinction substantially complicates the RANS prediction. Therefore, it poses a substantial challenge in reproducing the correct amount of extinction compared to the more precise LES and DNS approaches. Apart from the inaccuracies in the velocity field, the RngKE model is generally capable of reproducing the distribution of major species in flame F.

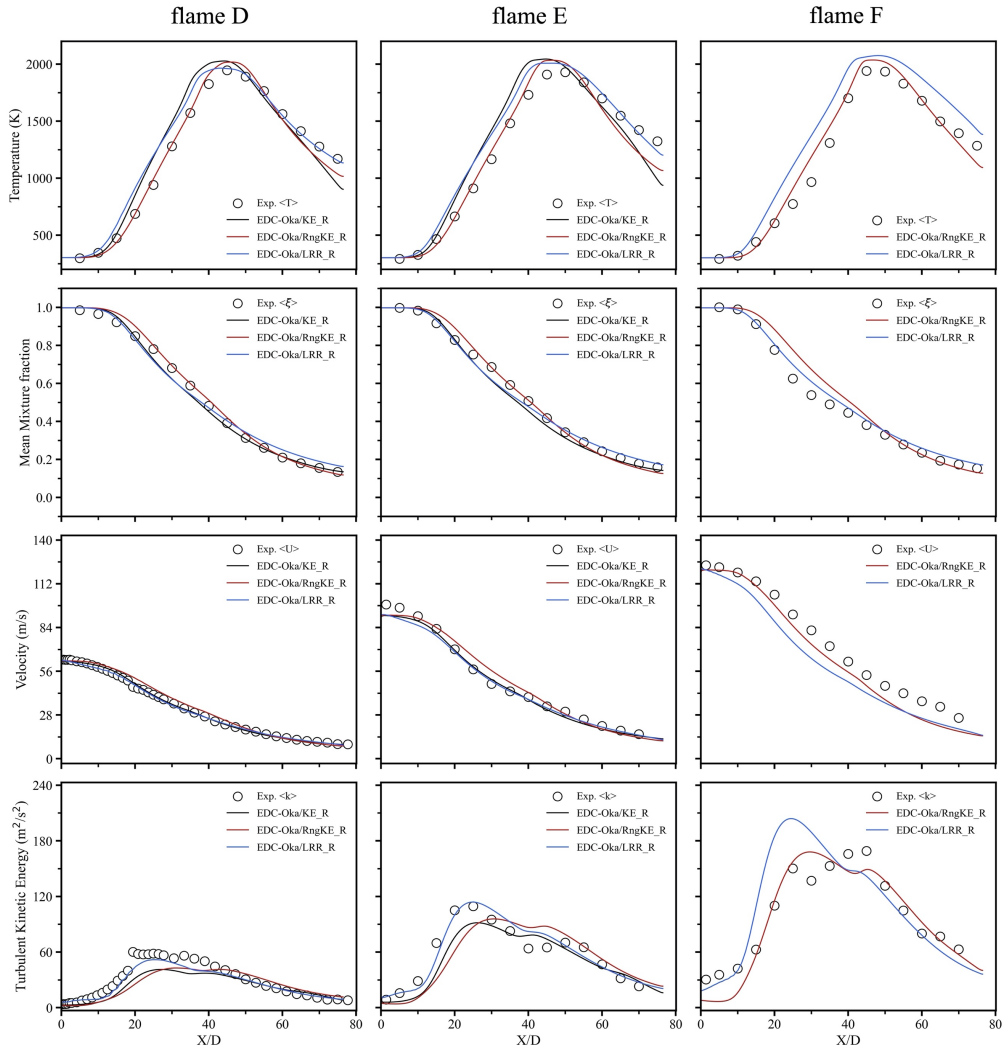

**Figure 5** Centerline profiles of (a) temperature (b) mean mixture value (c) normalized velocity (d) turbulent kinetic energy for flames D-F

Accurate simulation of minor species, particularly those affected by transient effects, remains a challenge in recent studies. As can be observed in Fig. 7, three turbulence models show an overprediction in peak values regarding the minor species, with such a phenomenon particularly pronounced in the prediction of NO. To calculate CO, employing the RngKE model results in the peak values for flames D and E being overestimated by approximately 0.34, and 0.35 times, respectively. The predictions for the LRR model display a marginal improvement over the RngKE model, at approximately 0.28 and 0.34 times. Saini et al. [31] attributed the intermediate CO discrepancy to the inadequate mixing time scale in the EDC combustion model. With respect to NO prediction, although all three turbulence models overestimate the NO prediction, the centerline NO profiles with the LRR model are observed to be closer to the experimental results. Compared to the RngKE model, the LRR model demonstrates enhanced prediction accuracy, overestimating flame D by 0.69 times and flame E by 1.13 times, as opposed to 1.41 and 1.86 times overestimation, respectively.

Fundamentally, the primary NO formation in gaseous combustion systems includes three mechanisms, namely, thermal NO, prompt NO, and fuel NO. Roomina and Bilger [32] studied flame D and reported that a skeletal mechanism including only thermal NO formation chemistry significantly underpredicted the NO mass fraction. They concluded that the predominant route for NO formation in the flame developing and combustion region with $X/D < 45$ is prompt NO. Overpredictions of NO in the present study can be related to the above theory due to the poor prediction of the prompt NO along the central axis. Further upstream near the flame front, where $X/D > 45$, thermal NO formation becomes the dominant process, and reasonable agreement for the LRR turbulence model is observed. The accuracy improvement in LRR turbulence model prediction can be linked to the precise prediction of peak flame temperature values, as shown in Fig. 5. Because thermal NO is one dominant source under most high-temperature circumstances, it is crucial to obtain precise high-temperature field estimates for accurate NO predictions.

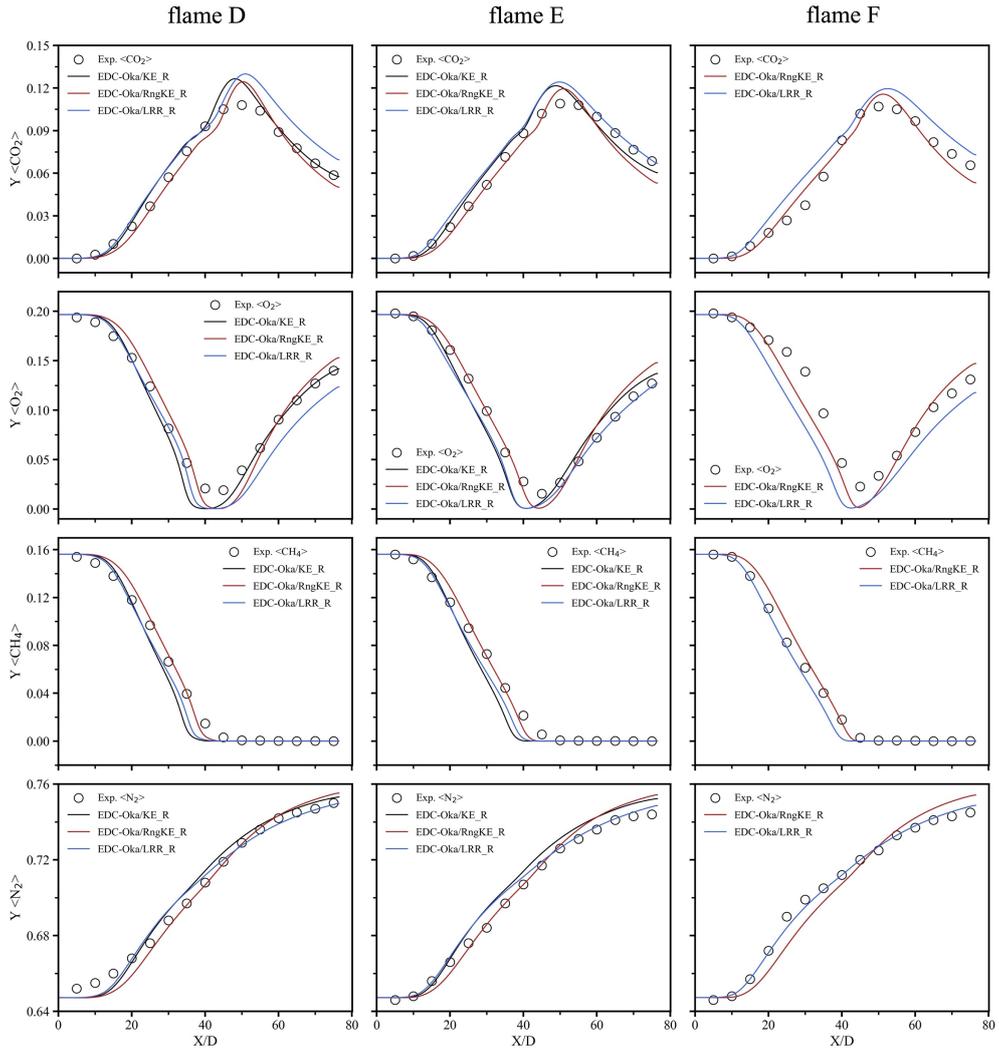

**Figure 6 Centerline profiles of (a) carbon dioxide (b) oxygen (c) methane (d) nitrogen for flames D-F**

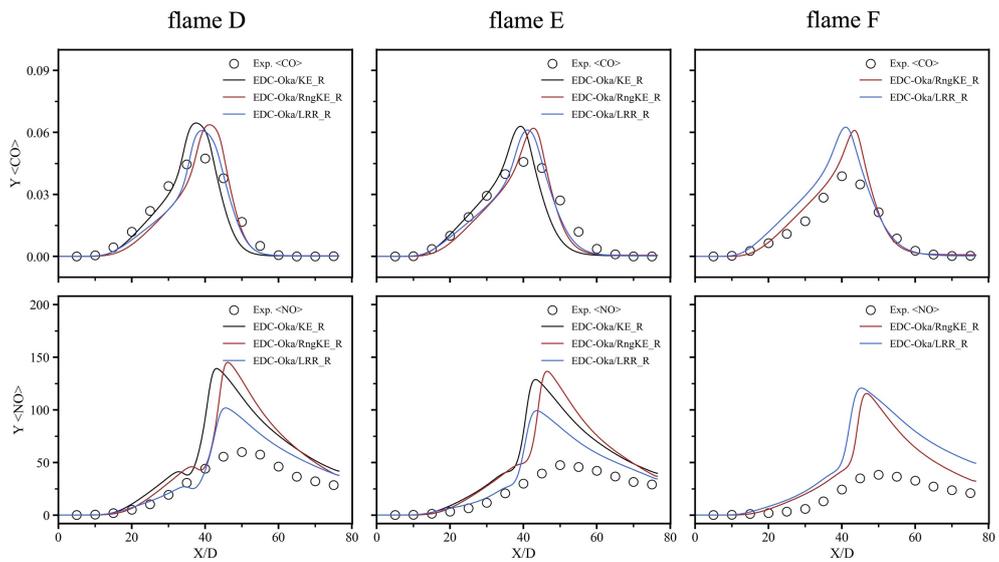

**Figure 7 Centerline profiles of (a) carbon monoxide (b) nitric oxide for flames D-F**

### 3.2.2 Radial distribution prediction

Figures 8 and 9 present the detailed radial profiles of temperature, mean mixture value, velocity, and turbulent kinetic energy for flame D. Take the RngKE model as an example, in the flame developing region $X/D \leq 7.5$, the predictions are basically in good agreement with the experimental data, demonstrating that the inlet boundary condition settings are appropriate. As the flame develops, the average axial velocity observed in the jet spreading is slightly overestimated at $X/D = 15$, resulting in the prediction of large temperature values. Similar overestimations are observed in other RANS-related studies applying the Standard $k$–$\varepsilon$ [31], Realizable $k$–$\varepsilon$ [33], and Reynolds stress model [34] respectively. At the flame front when $X/D = 45$ and 60, although temperature field overestimations persist, the prediction of jet spreading becomes accurate, and turbulent kinetic energy profiles are reasonably well predicted at both locations. Collectively, the aforementioned findings indicate that the RngKE model is capable of providing reasonably accurate predictions for flame D.

Regarding the predictions of the other two KE and LRR turbulence models, differences in jet spreading exhibit little deviation from the RngKE predictions, and experimental results are well reproduced at most radial locations. However, the problem of the temperature field overestimation is more pronounced, especially at the axial positions of $X/D = 7.5$, 15, and 30, where the predictions for the axial temperature distribution are significantly higher than the experimental values. One explanation for this deviation is the selection of empirical coefficients in the LRR model, as it typically requires variable case-specific empirical data. Differences in these empirical coefficients could constrain the prediction accuracy and impose a limitation on the practical application of the LRR model [20].

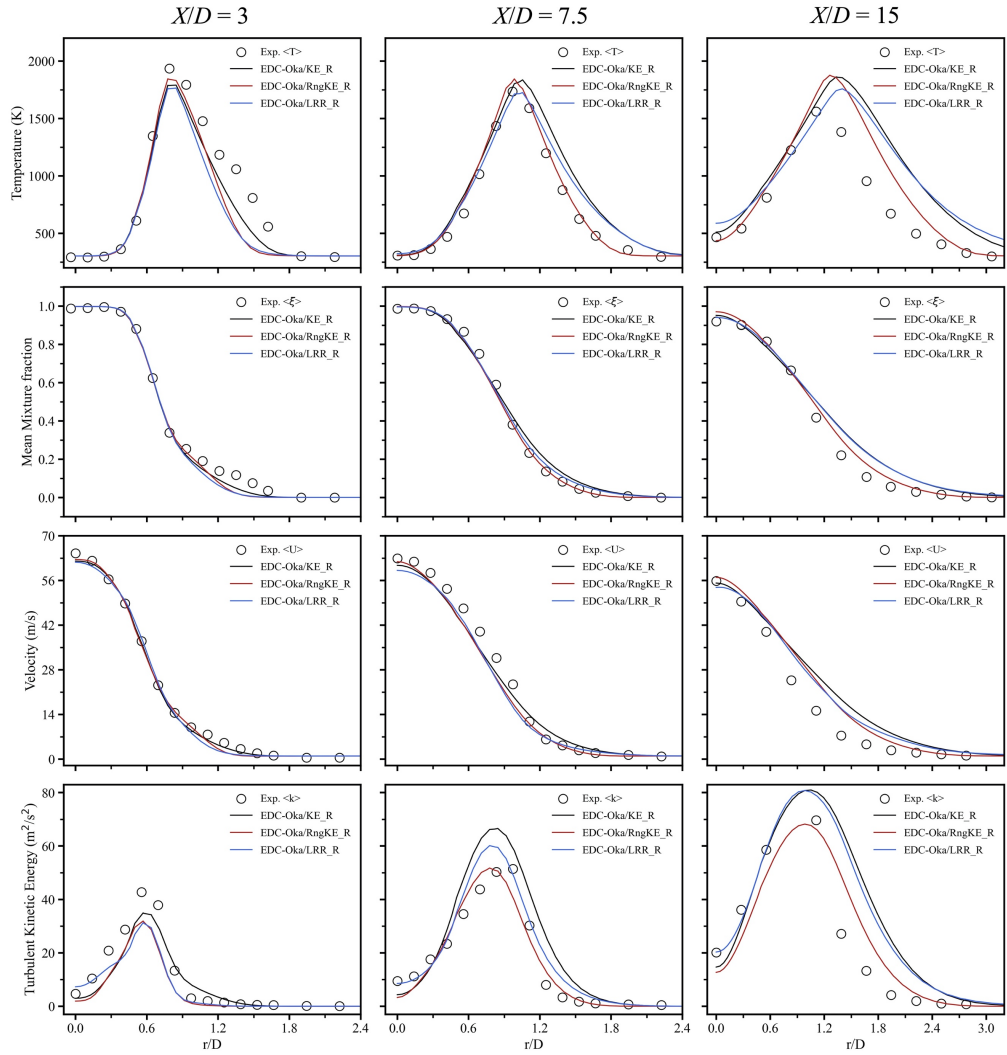

**Figure 8 Radial profiles of (a) temperature (b) mean mixture value (c) normalized velocity (d) turbulent kinetic energy at *X*/*D* = 3 (left), *X*/*D* = 7.5 (middle) and *X*/*D* = 15 (right) for flame D**

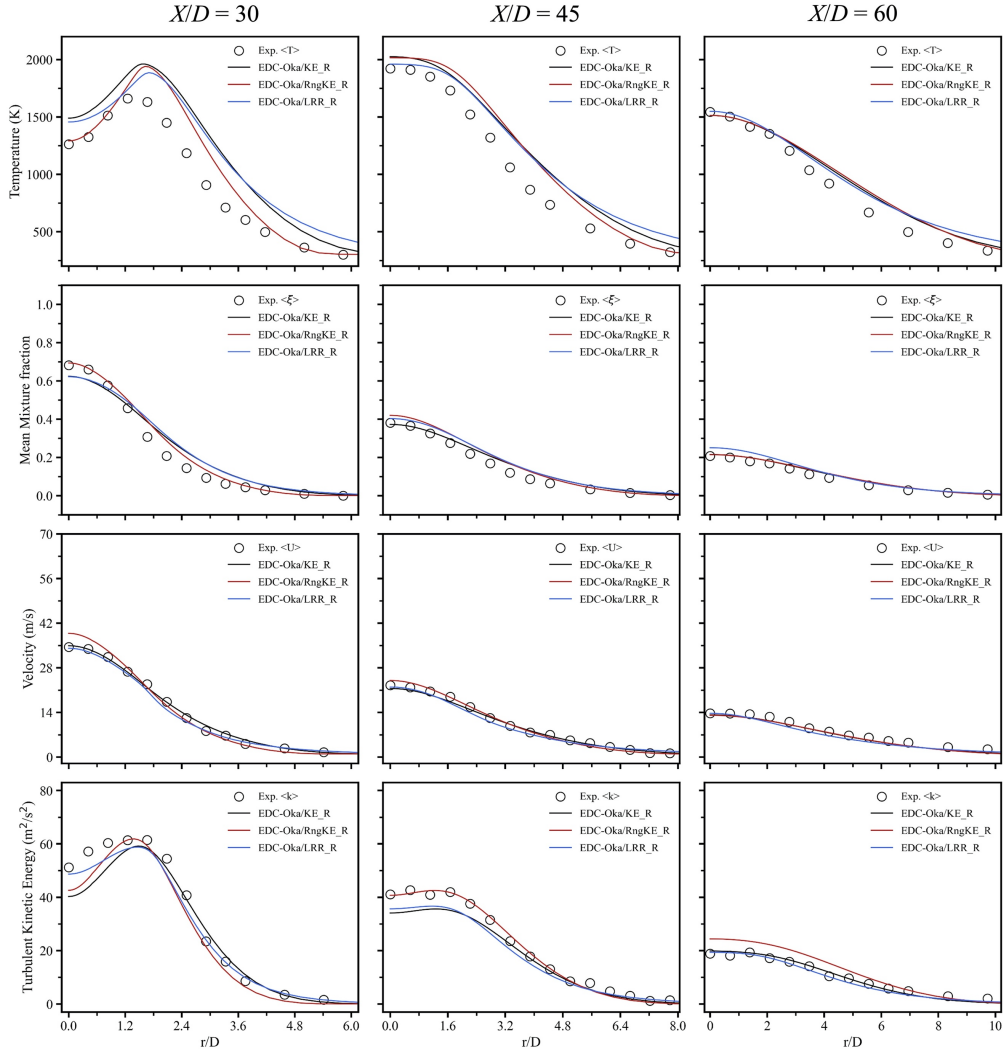

**Figure 9** Radial profiles of (a) temperature (b) mean mixture value (c) normalized velocity (d) turbulent kinetic energy at $X/D = 30$ (left), $X/D = 45$ (middle), and $X/D = 60$ (right) for flame D

Figure 10 presents the radial profiles of the NO for flame D to flame F at three different axial locations. In terms of radial distribution, the LRR model also demonstrates higher accuracy in NO prediction, while results based on the RngKE and KE model are considerably overestimated. In various jet burning studies concerning hydrocarbon or decarbonized fuels, controlling fuel and oxidizer inlet flow velocities is a prevalent strategy, resulting in a rich fuel region that suppresses NO formation [2]. This technique effectively minimizes the NO formation in the majority of ammonia-related combustion furnaces studies. The three turbulence models applied in the present study reproduce that phenomenon well through the flow velocity transition (from flame D to flame F), except for a slight overestimation of flame F. This discrepancy could be attributed to the limited precision in simulating severe local extinction and consequently increased overprediction of mean temperatures within that region, which in turn promotes the generation of additional thermal NO.

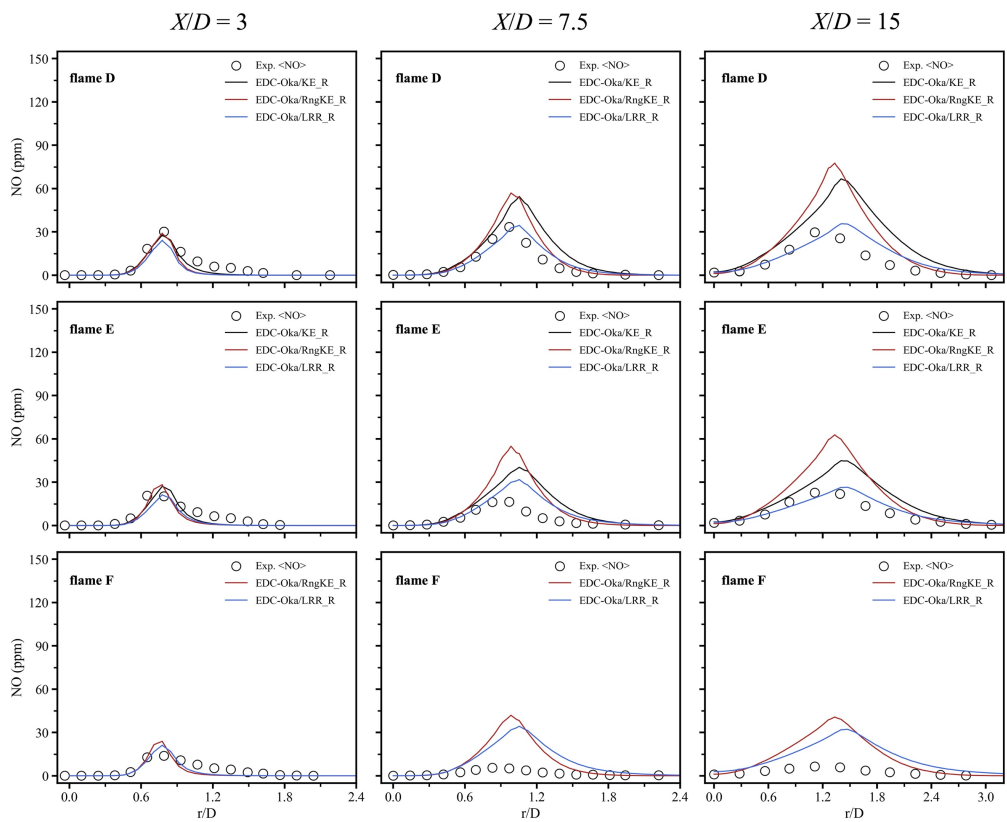

**Figure 10 Radial profiles of nitric oxide (NO) for flame D to flame F at $X/D$ = 3 (left), $X/D$ = 7.5 (middle), and $X/D$ = 15 (right)**

**3.3 LRR model constant assessment**

It is known that several variables can affect the prediction accuracy in CFD calculations, including the reaction mechanism [18], combustion model [19, 35, 36], and as previously discussed, inlet boundary conditions and turbulence models. In the case of turbulence models, the optimization of turbulence model constants, plays a crucial role in achieving accurate predictions, especially to the $C_{\varepsilon1}$ (related to the dissipation rate). One underlying reason is that the default constant of the turbulence model fails to provide accurate predictions of the spreading rate and dissipation rate under different combustion conditions. Currently, many researchers have put forth recommended values for the $C_{\varepsilon1}$ within the KE turbulence model according to their customized solver [37, 38]. However, to the best of the authors' knowledge, turbulence model constants of the LRR turbulence model have rarely been discussed. In this section, different turbulence model constants for the LRR model are examined, based on the solver applied in the present study to determine the optimal prediction solution for the round jet flow.

Figure 11 displays a comparison of the mean temperature distribution for flame D under different model constants using the LRR model. The predicted scalar data for the central axis, $X/D = 7.5$ (flame developing region), and $X/D = 60$ (flame front) are presented in Fig. 12. Evidently, as the $C_{\varepsilon1}$ gradually increases from the default value (1.44) to 1.5, the turbulence dissipation rate is enhanced, and more turbulence is converted to heat through molecular viscosity, which is particularly evident at the flame front. In the flame-developing region, different cases all over-predict the temperature field, with a pronounced effect when $C_{\varepsilon1}$ is equal to 1.44. At the flame front, an increase in the $C_{\varepsilon1}$ significantly improves the energy transfer efficiency, resulting in a higher temperature distribution. When $C_{\varepsilon1}$ equals 1.48, predictions correspond well with experimental data, while the temperature field becomes overestimated as the value increases to 1.5. However, for the velocity field, an opposite trend was observed. As $C_{\varepsilon 1}$ changes from 1.44 to 1.5, the dissipation rate increases, accompanied by an enhanced conversion of turbulence into heat. This conversion consistently acts to reduce the turbulent kinetic energy, resulting in a higher deviation of the turbulent kinetic energy from the experimental value at $C_{\varepsilon 1}$ = 1.5.

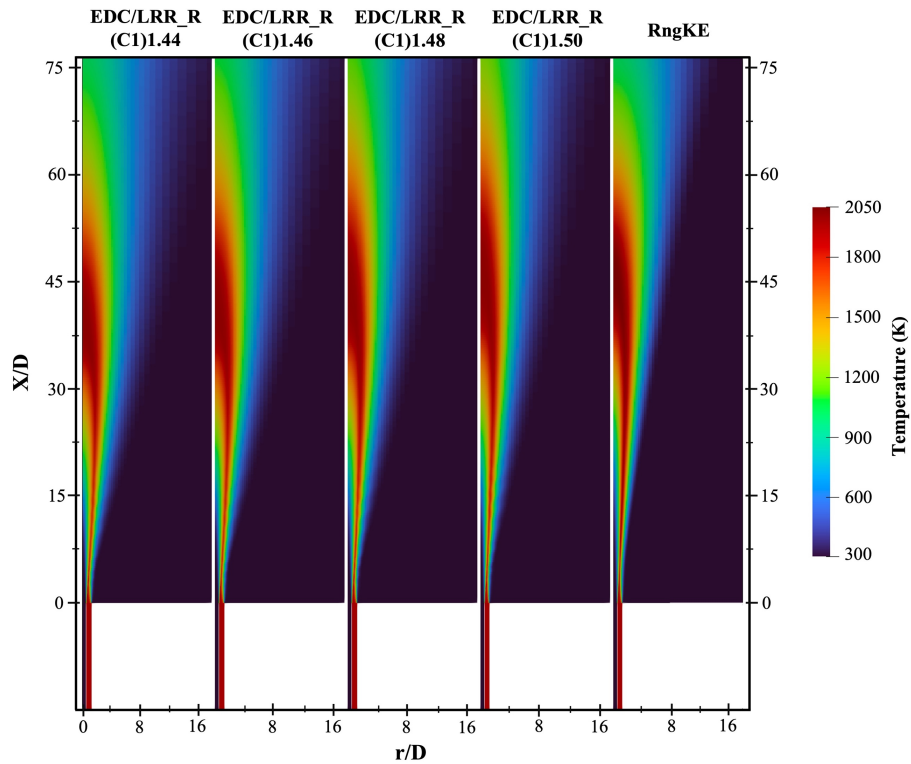

**Figure 11 Mean temperature distribution of the flame D under different model constants of the LRR model**

In summary, based on the customized solver applied in the present study, adjusting the $C_{\varepsilon 1}$ from 1.44 to 1.48 in the LRR model achieves better prediction results. A comprehensive analysis of the quantitative prediction performance concerning different model constants will be elaborated upon in Section 3.4. However, notably, although accurate predictions can be obtained after adjusting the $C_{\varepsilon 1}$, the adjustment of the model constants is of limited value and the notion of generality is lost. As Pope [37] stated that adjusting the model constant value might yield the desired predictions, but in doing so, will simultaneously lose the sense of generality.

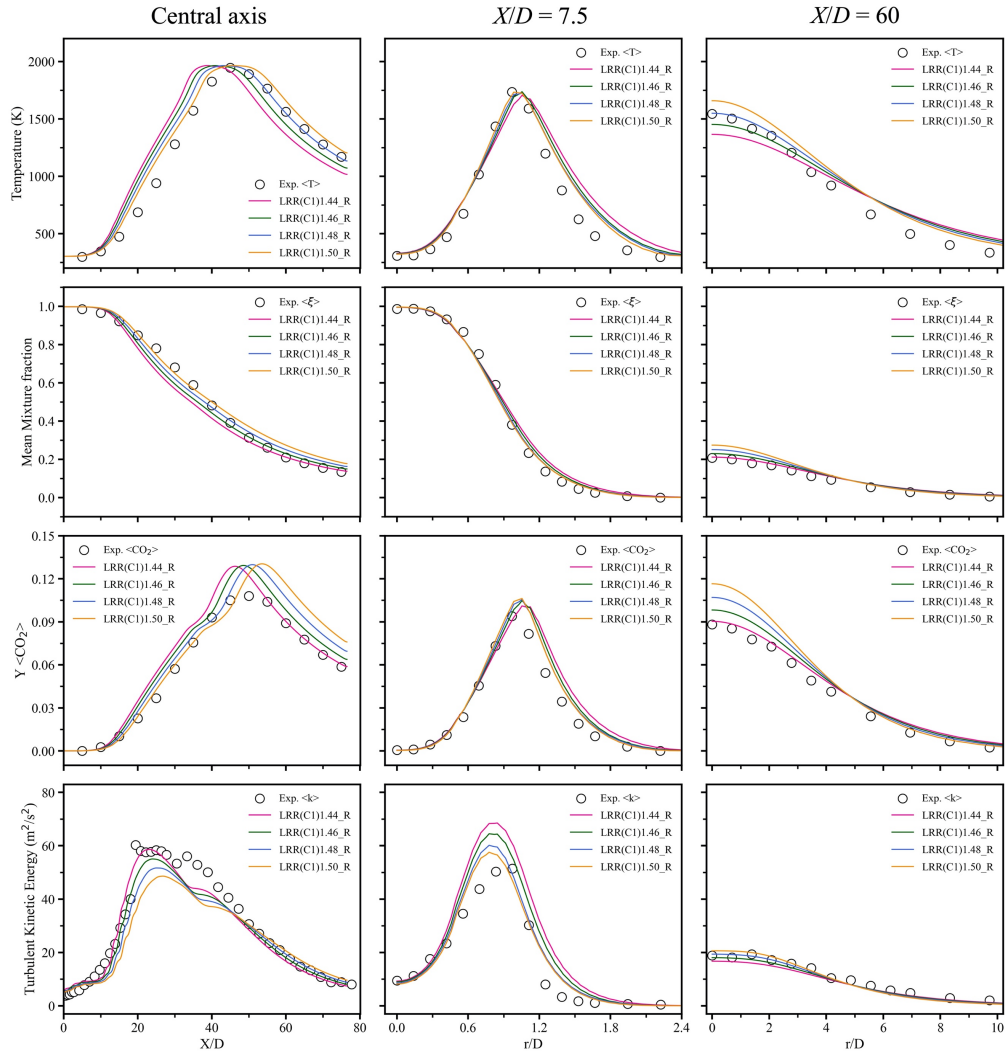

**Figure 12 Influence of LRR model constant ($C_{\varepsilon 1}$) to the prediction results of flame D**

## 3.4 Evaluation of various turbulence-combustion models

The assessment of generality and accuracy of various turbulence models has been a crucial topic in combustion predictions, especially when the results cannot be visually discerned. Consequently, the present section quantifies the prediction results derived from an array of turbulence and combustion models to demonstrate their prediction performance. Owing to the high quality of results from flame D, the analysis will predominantly focus on the prediction performance of flame D.

The root-mean-square error (RMSE) serves as an effective method for assessing prediction accuracy and is frequently employed as a performance metric [39]. The RMSE can be regarded as an uncertainty metric when different calculation models are used as a predictor, and its value can be expressed by Eq. (15)

$$RMSE = \frac{1}{\widetilde{\Omega}} \sqrt{\frac{1}{n} \sum_{i=1}^{n} [f(x_i) - y_i]^2} \qquad (15)$$

where $n$ is the number of experimental points, $f(x_i)$ and $y_i$ are the predicted and experimental results, respectively. The expression of $\widetilde{\Omega}$ can be denoted in Eq. (16)

$$\widetilde{\Omega} = \sqrt{\frac{1}{n} \sum_{i=1}^{n} y_i^2} \qquad (16)$$

where $\widetilde{\Omega}$ is the normalization constant, which can be obtained from the experimental results. The calculated RMSE values for flame D of various cases along the central axis is shown in Table 2. In light of the synthesis of prediction results across different cases, three primary conclusions can be drawn. Firstly, the RngKE and LRR turbulence models yield satisfactory predictions. Specifically, the RngKE model provides accurate predictions of the temperature field and major species ($CO_2$, $CH_4$, $O_2$). While the LRR model, better reproduces turbulence effects, thus exhibiting higher accuracy in terms of velocity and turbulent kinetic energy. Additionally, with the LRR model constant $C_{\varepsilon 1}$ set to 1.48, the RMSE value is most reasonable, validating the accuracy of the model constant selection. Secondly, for the NO predictions, although the RMSE values for all three types of models display significant deviations from the experimental values, the LRR model offers superior prediction accuracy compared to the other two models. One explanation pertains to the LRR model accurately predicting the high-temperature field at the flame front, which reduces the thermal NO production. Thirdly, the present study compares differences between the Okafor detailed reaction mechanism and the GRI-3.0. Despite slightly better predictions for the GRI-3.0, the RMSE values obtained under GRI-3.0 and the Okafor detailed mechanism are approximately identical.

**Table 2 RMSE comparison of various calculation cases for major species of flame D along the central axis**

| Case | T | $CO_2$ | F | U | k | $CH_4$ | $O_2$ | NO | $N_2$ |
|---|---|---|---|---|---|---|---|---|---|
| EDC/RngKE_R | 4.60 | 8.98 | 4.12 | 5.97 | 40.13 | 7.53 | 10.21 | 92.46 | 0.68 |
| EDC/RngKE_R (*GRI 3.0*) | 4.71 | 8.92 | 4.03 | 5.87 | 39.74 | 7.41 | 10.13 | 92.01 | 0.66 |
| EDC/RngKE | 8.47 | 12.51 | 6.36 | 6.91 | 47.31 | 6.94 | 12.41 | 159.15 | 0.78 |
| EDC/LRR_R_(C1)1.44 | 16.22 | 15.25 | 9.34 | 5.63 | 11.39 | 18.46 | 12.25 | 40.72 | 1.03 |
| EDC/LRR_R_(C1)1.46 | 12.3 | 14.28 | 6.91 | 3.54 | 12.68 | 14.43 | 11.14 | 39.34 | 0.74 |
| EDC/LRR_R_(C1)1.48 | 8.61 | 16.13 | 5.72 | 2.01 | 16.92 | 9.31 | 13.87 | 48.86 | 0.56 |
| EDC/LRR_R_(C1)1.50 | 6.98 | 20.54 | 6.76 | 2.53 | 23.74 | 5.56 | 19.98 | 56.09 | 0.64 |
| EDC/KE_R | 10.88 | 9.68 | 4.65 | 2.13 | 32.99 | 13.13 | 9.78 | 89.35 | 0.66 |
| PaSR/RngKE | 9.49 | 15.02 | 7.39 | 7.48 | 50.39 | 7.98 | 13.26 | 184.54 | 0.93 |

For the third point, it should be mentioned that, usually, reaction mechanisms have the most pronounced effects on emission predictions, and the NO is especially sensitive to different reaction mechanisms. It is unclear as to why the difference is not prominent. Given that the Okafor detailed mechanism was built based on GRI-3.0 and is intended for the prediction of ammonia combustion [6]. The comparable results to GRI-3.0 validates its prediction accuracy. Therefore, more attention can be given to the Okafor detailed mechanism in future ammonia-related studies.

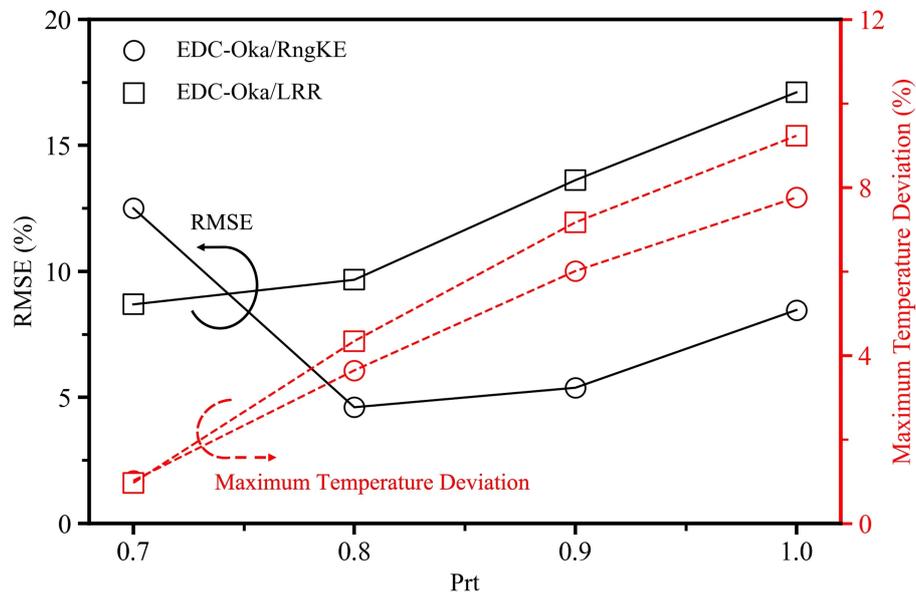

**Figure 13 Influence of turbulent Prandtl number on the temperature field for flame D**

In most CFD modeling studies, an accurate turbulent Prandtl number setting poses a considerable impact on predictions. In other words, the turbulent Prandtl number is a concept pertinent to turbulent heat transfer, and thus, it significantly affects the temperature field. Figure 13 illustrates the influence of turbulent Prandtl numbers on the temperature field. As can be observed, optimal turbulent Prandtl numbers vary when employing different turbulence models. Low RMSE values and low-temperature deviations are achieved when the turbulent Prandtl number is set to 0.72 and 0.82 for the LRR and RngKE models, respectively. Nevertheless, the influence of turbulent Prandtl number in combustion prediction research often goes beyond that. Typically, the combustion models assume that the turbulent field transports all scalars, such as $CH_4$, and $CO_2$, in a similar way as heat [28]. Moreover, reactions of thermal NO are highly temperature dependent, rendering these species highly sensitive to the turbulent

Prandtl number. Combining Figs. 7, 10, 13, and Table 2, the above conclusion explains why the LRR model outperforms the other two turbulence models in terms of NO prediction, which is attributed to the lowered turbulent Prandtl number setting.

## 4. Conclusions

OpenFOAM has been widely used in the CFD community, but its original performance concerning computational efficiency and accuracy in simulating reacting flows has not been satisfactory. The present study introduces a fast and accurate customized solver, developed on the OpenFOAM platform, specifically designed for ammonia industrial applications.

First, with regard to augmenting computational efficiency, the present study developed a combined acceleration strategy. These improvements include: (1) a sparse analytical Jacobian approach utilizing the SpeedCHEM chemistry library was implemented to optimize chemistry ODE solution routines; (2) the DLB code was employed to evenly redistribute the computational load for chemistry among multiple processes; (3) the OpenMP method was introduced to enhance parallel computing efficiency; (4) the LTS scheme was integrated to maximize the individual time step for each computational cell. With the application of such a computational strategy, a minimum of a 31-fold speed-up was achieved when compared to the standard approach.

In the following, supplementary optimizations were performed on the governing equations, turbulence models, combustion models, and radiation models within the native reactingFoam solver on the OpenFOAM platform. To assess the prediction accuracy of the customized solver, the present investigation employed the RANS method to validate the temperature fields, velocity fields, and species distributions of three distinct partially turbulent premixed flames: Sandia Flames D, E, and F. Through a comprehensive comparative analysis involving various turbulence models, turbulent Prandtl numbers, and model constants, optimal numerical parameters were identified for various conditions. Overall, the prediction results demonstrate that all three turbulence models, after the appropriate selection of wall functions, are capable of reasonably reproducing the experimental outcomes. Regarding the flame D calculation, the LRR model achieves better prediction results when the model constant and turbulent Prandtl number are set to 1.48 and 0.72, respectively. In contrast, increasing the turbulent Prandtl number by 0.1 for RngKE and KE models yields improved results. The RngKE turbulence model demonstrates higher accuracy in the temperature field and major species predictions, while the LRR model exhibits superior precision in velocity field predictions. Concerning the NO prediction, the LRR model provides noticeably better results than the other two models, as it predicts the peak temperature with improved accuracy, which reduces the formation of thermal NO. The investigation particularly focused on the impact of the turbulent Prandtl number on the NO prediction, revealing that increasing the turbulent Prandtl number in scenarios with increased inlet jet velocities yielded improved results. The customized solver developed in the present study has been verified and can be implemented in subsequent industrial ammonia combustion scenarios.

## Acknowledgement

The authors would like to thank Editage (www.editage.com) for English language editing. Additionally, the first author gratefully acknowledges financial support from China Scholarship Council (Grant No. 202106250035).